\documentstyle[psfig,epsf]{article}
\def \lsim
{\raisebox{-3pt}{$\>\stackrel{<}{\scriptstyle\sim}\>$}}

\newcommand{\msbar}{$\overline{\mbox{MS}}$ }

\begin{document}
\begin{flushright}
CERN-PH-TH/2005-016\\
February 2005
\end{flushright}

\begin{center}
{\Large \bf Progress in bottom-quark fragmentation} 
\footnote{Talk given at 
18th International Workshop on High Energy Physics 
and Quantum Field Theory (QFTHEP 2004), 
St.Petersburg, Russia, 17--23 June 2004.}\\
\vspace{4mm}

G. Corcella\\
Department of Physics, CERN\\ 
Theory Division\\
CH-1211 Geneva 23, Switzerland 
\end{center}

\begin{abstract}
\noindent
I review recent progress on bottom-quark fragmentation in 
$e^+e^-$ annihilation, top-quark decay and $H\to b\bar b$ processes.
I discuss the implementation of collinear and soft resummation, and the
inclusion of non-perturbative information taken from LEP and SLD data.
\end{abstract}

Heavy-quark physics is currently one of the major fields of investigation 
in experimental and theoretical particle physics.
In order to investigate heavy-quark phenomenology, fixed-order calculations
are reliable as long as one considers inclusive observables, such as
total cross sections. For less inclusive quantities, one needs to resum 
large contributions, which correspond to collinear or soft parton radiation.

In this talk I will investigate bottom-quark production in 
$e^+e^-\to b\bar b$ processes, top-quark decay $t\to bW$, and 
Standard Model Higgs decay $H\to b\bar b$.
In particular, I wish to study the energy distribution of $b$ quarks
and $b$-flavoured hadrons in such processes.
The results will be expressed in terms of 
the $b$ normalized energy fraction $x_b$, which is given, e.g.
in Higgs decay $H\to b\bar b(g)$, by:
\footnote{In Eq.~(\ref{xb}) powers of $m_b/m_H$ are neglected.
In top decay we shall have an extra term to account for the $W$ 
mass \cite{cormit}.}
\begin{equation}
x_b={{2p_b\cdot p_H}\over{m_H^2}}.
\label{xb}
\end{equation}
The differential distribution for the production of a
massive $b$ quark at next-to-leading
order (NLO) in the strong coupling
constant $\alpha_S$ is given by the following general result:
\begin{equation}
{1\over{\Gamma_0}}{{d\Gamma}\over {dx_b}}=\delta(1-x_b)
+{{\alpha_S(\mu)}\over{2\pi}}
\left[P_{qq}(x_b)\ln{{m_H^2}\over{m_b^2}}+A(x_b)\right] +
{\cal O} \left( {{m_b^2}\over{m_H^2}}\right)^p  .
\label{massb}
\end{equation}
In Eq.~(\ref{massb}) $\Gamma_0$ is the width of the Born process,
$\mu$ is the renormalization scale, $A(x_b)$ is a function independent
of $m_b$, $p\geq 1$,
$P_{qq}(x_b)$ is the Altarelli--Parisi splitting function:
\begin{equation}
P_{qq}(x_b)=C_F\left( {{1+x_b^2}\over {1-x_b}}\right)_+.
\end{equation}
Equation~(\ref{massb}) is formally equal to the bottom energy spectrum
in top decay or $e^+e^-$ processes: one will just have to replace 
$m_H$ with the top mass or the centre-of-mass energy $\sqrt{s}$ of the
$e^+e^-$ collision. Function $A(x_b)$ is instead process-dependent.

Equation~(\ref{massb}) presents a large mass logarithm 
$\sim\alpha_S\ln(m_H^2/m_b^2)$,
which needs to be resummed to all orders to improve the prediction.
To achieve this goal,
we can follow the approach of perturbative fragmentation functions 
\cite{mele}, which expresses the energy spectrum of a heavy quark as the 
convolution of a coefficient function, describing the emission of a 
massless parton, and a perturbative fragmentation function $D(m_b,\mu_F)$,
associated with the transition of a massless parton into a massive
quark:
\begin{eqnarray}
{1\over {\Gamma_0}} {{d\Gamma_b}\over{dx_b}} (x_b,m_H,m_b) &=&
\sum_i\int_{x_b}^1
{{{dz}\over z}\left[{1\over{\Gamma_0}}
{{d\hat\Gamma_i}\over {dz}}(z,m_H,\mu,\mu_F)
\right]^{\overline{\mathrm{MS}}}
D_i^{\overline{\mathrm{MS}}}\left({x_b\over z},\mu_F,m_b \right)} \nonumber \\
&+& {\cal O}\left((m_b/m_H)^p\right) \; .
\label{pff}
\end{eqnarray}
In Eq.~(\ref{pff}), $d\hat\Gamma_i /dz$ is the differential width for the 
production of a massless parton $i$ in Higgs decay with 
an energy fraction $z$;
$D_i(x,\mu_F,m_b)$ is the perturbative
fragmentation function for a parton $i$ to fragment
into a massive $b$ quark, $\mu_F$ is the factorization scale.
Neglecting $g\to b\bar b$ splitting, $i=b$ on the right-hand side
of Eq.~(\ref{pff}). 
The coefficient function for $H\to b\bar b$ processes has been computed in
\cite{cor} and reads, in the \msbar factorization scheme:
\begin{eqnarray}
\left[{1\over{\Gamma_0}} {{d\hat\Gamma_b}\over{dz}} (z,m_H,\mu,\mu_F)
\right]^{\overline{\mathrm{MS}}}
&=&
\delta(1-z)+{{\alpha_S (\mu)C_F}\over{2\pi}}
\left[\left({{1+z^2}\over{1-z}}\right)_+\ln{{m_H^2}\over{\mu_F^2}} 
\right.\nonumber\\
&+&
\left( {2\over 3} \pi^2
+ {3\over 2}\right) \delta(1-z)
+1-z-{3\over 2} {{z^2}\over{(1-z)_+}}\nonumber\\
&-&(1+z)[\ln(1-z)+2\ln z]
+6{{\ln z}\over{(1-z)_+}}\nonumber\\
&-&2{{\ln z}\over {1-z}}
+2\left.\left({{\ln(1-z)}\over{1-z}}\right)_+\right].
\label{coeffx}
\end{eqnarray}
As pointed out in Ref.~\cite{cor}, in the calculation of Eq.(\ref{coeffx}),
using the \msbar-renormalized Yukawa coupling $\bar y_b(m_b)$ 
for the $Hb\bar b$ vertex turned out to be essential.
The \msbar NLO coefficient function for $e^+e^-$ annihilation was 
calculated in \cite{mele}, for top decay in \cite{cormit}.

The perturbative fragmentation function follows the 
Dokshitzer--Gribov--Altarelli--Parisi (DGLAP) evolution equations
\cite{ap,dgl}. Its value at a given scale $\mu_F$ can be obtained once
an initial condition is given. In \cite{mele} the initial condition
$D_b^{\rm ini}(x_b,\mu_{0F},m_b)$ was calculated and its process-independence 
was established on more general grounds in \cite{cc}.
It is given at NLO by:
\begin{equation}
D_b^{\rm ini}(x_b,\mu_{0F},m_b)=\delta(1-x_b)+
{{\alpha_S(\mu_0^2)C_F}\over{2\pi}}
\left[{{1+x_b^2}\over{1-x_b}}\left(\ln {{\mu_{0F}^2}\over{m_b^2}}-
2\ln (1-x_b)-1\right)\right]_+.
\label{dbb}
\end{equation}
The NNLO initial condition was calculated in Refs.~\cite{alex1,alex2}.

As discussed in \cite{mele}, solving the DGLAP equations for an evolution
from $\mu_{0F}$ to $\mu_F$, with a NLO kernel, allows one to resum leading (LL)
$\alpha_S^n\ln^n(\mu_F^2/\mu_{0F}^2)$ and next-to-leading (NLL) 
$\alpha_S^n \ln^{n-1}(\mu_F^2/\mu_{0F}^2)$ logarithms (collinear
resummation).
The explicit expression for the solution of the DGLAP equations can be found,
e.g., in Ref.~\cite{mele}.

In particular, in $H\to b\bar b$ processes,
for an evolution from $\mu_{0F}\simeq m_b$ to $\mu_F\simeq m_H$, the large 
logarithms $\ln(m_H^2/m_b^2)$ appearing in the massive calculation
(\ref{massb}) are resummed with NLL accuracy.
Likewise, since the perturbative 
fragmentation approach and the collinear resummation are process-independent, 
we can resum, following the same procedure, NLL logarithms $\ln(s/m_b^2)$ and 
$\ln(m_t^2/m_b^2)$ in the massive $b$ spectrum in $e^+e^-$ annihilation 
\cite{mele} and top decay \cite{cormit}.

Furthermore, both the coefficient functions in Eq.~(\ref{coeffx})
and Refs.~\cite{cormit,cc} and the
initial condition of the perturbative fragmentation function (\ref{dbb}) 
present terms that become large once the $b$-quark energy fraction $x_b$ 
approaches 1, which corresponds to soft-gluon radiation.
Soft contributions to the initial condition are process-independent and 
were resummed in \cite{cc} in the NLL approximation.
NLL soft-gluon resummation 
in the coefficient function is instead process-dependent and it was implemented
in $e^+e^-$ annihilation \cite{cc}, top decay \cite{ccm} and
$H\to b\bar b$ \cite{cor} processes. For the sake of brevity,
I do not report here the expression for the soft-resummed initial condition
and coefficient functions.

\begin{figure}
\centerline{\psfig{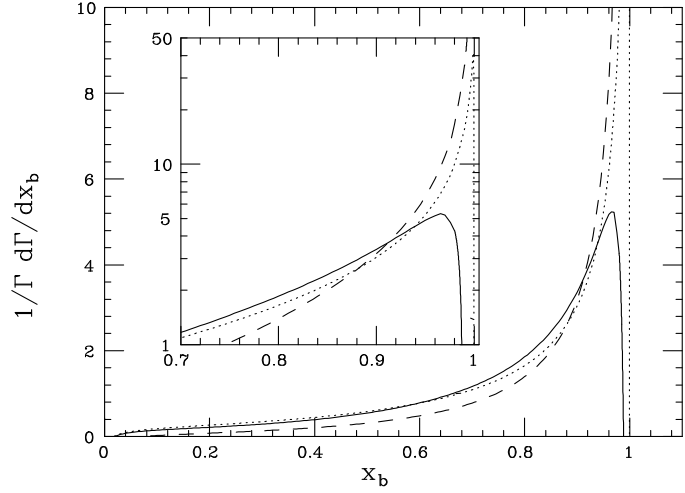}}
\caption{$b$ spectrum in Higgs decay according to
the NLO massive calculation (dashed), including NLL collinear resummation
(dotted), and both NLL collinear and soft resummations (solid).   
In the inset, the same curves are shown at large $x_b$, on a logarithmic
scale.} 
\label{hxb}
\end{figure}

In Fig.~\ref{hxb}, the bottom energy distribution in Higgs decay is
given; the result is qualitatively similar to what can be found
in $e^+e^-$ processes or top decay.
I have plotted the NLO massive calculation (dashed), 
including NLL collinear resummation (dotted), and with both NLL 
collinear and soft resummations (solid). I have set 
$\mu=\mu_F=m_H$, $\mu_0=\mu_{0F}=m_b$, $m_H=120$~GeV, $m_b=5$~GeV.
The fixed-order calculation lies below the resummed ones and 
grows as $x_b\to 1$, since $\sim 1/(1-x_b)$;
the collinear-resummed spectrum exhibits instead a sharp peak at large $x_b$.
Soft resummation is relevant at $x_b>0.6$:
the distribution is further smoothed and shows the Sudakov peak at 
$x_b\simeq 0.97$.
References~\cite{cormit,cor,cc} 
show that, after collinear and soft logarithms are
resummed, the $x_b$ distribution exhibits very mild dependence on factorization
and renormalization scales, which corresponds to a reduction of the theoretical
uncertainty on the prediction.

\begin{figure}
\centerline{\psfig{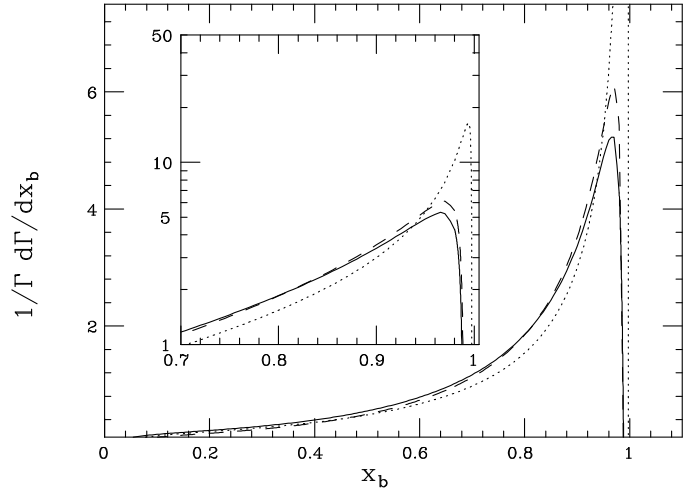}}
\caption{$b$ spectrum in Higgs decay, $e^+e^-$ annihilation 
at $\sqrt{s}=91.2$~GeV (dashes),
top decay (dots). All curves include NLL collinear and soft resummation.}
\label{hzt}
\end{figure}

It is interesting to compare the $b$ spectrum in the 
considered three processes.
In Fig.~\ref{hzt} we plot the $x_b$ distributions in $H\to b\bar b$,
$e^+e^-\to b\bar b$ at $\sqrt{s}=m_Z$, i.e. LEP I and SLD 
centre-of-mass energy, and $t\to bW$. 
The $b$ spectrum in $H$ decay is the highest at small $x_b$ and the
lowest at large $x_b$;
in top decay it is shifted toward large $x_b$ and 
sharply peaked very close to 1.
The $e^+e^-\to b\bar b$ prediction is similar to the one yielded 
by Higgs decay, although some difference is still visible around the
Sudakov peak.
It was shown in \cite{cor} that setting $\sqrt{s}=m_H$ makes the predictions
in $H\to b\bar b$ and $e^+e^-$ annihilation more similar still.
The difference noticed between the $b$ spectra in Higgs and top decay can be 
experimentally relevant when distinguishing jets initiated by
$b$ quarks in the process $pp\to t\bar t H$, followed by $H\to b\bar b$.
In fact, this is one of the search channels of the Standard Model Higgs boson 
at the LHC, for a Higgs mass $m_H\lsim 135$~GeV \cite{denegri}.

I would like to present results for $b$-flavoured $B$ hadron production
in the investigated processes. 
The $B$ spectrum will be given by the convolution of the $b$-energy 
distribution with a non-perturbative fragmentation function associated with
the hadronization.
As in \cite{cormit,cor,ccm}, I consider a power law with two tunable
parameters
\begin{equation}
D^{np}(x;\alpha,\beta)={1\over{B(\beta +1,\alpha +1)}}(1-x)^\alpha x^\beta,
\label{ab}
\end{equation}
the model of Kartvelishvili et al. \cite{kart}
\begin{equation}
D^{np}(x;\delta)=(1+\delta)(2+\delta) (1-x) x^\delta,
\label{kk}
\end{equation}
and the Peterson model \cite{peterson}
\begin{equation}
D^{np}(x;\epsilon)={A\over {x[1-1/x-\epsilon/(1-x)]^2}}.
\label{peter}
\end{equation}
In Eq.~(\ref{ab}), $B(x,y)$ is the Euler beta function; in
(\ref{peter}) $A$ is a normalization constant.

The parameters $\alpha$, $\beta$, $\delta$ and $\epsilon$ can be obtained 
after fitting models (\ref{ab})--(\ref{peter}) to $x_B$ data in
$e^+e^-$ collisions, where $x_B$ is the normalized $B$ energy fraction.
In \cite{cor,ifae}, data from ALEPH \cite{aleph}
and SLD \cite{sld} Collaborations were considered.
In \cite{ifae} it was reported that if we fit independently
the non-perturbative models to ALEPH and SLD, we get pretty different
best-fit parameters.
Unlike ALEPH, which detected $B$ mesons,
SLD was able to reconstruct some $B$ baryons as well;
however it is a very small
fraction of the whole sample and that may not be the reason for the 
discrepancies discussed in \cite{ifae}.

In \cite{cor} a combined fit to ALEPH and SLD was performed instead, as if all
data came from the same sample. This approach is used here too.
In the fits, the correlations among data points are neglected,
and the considered data are in the range $0.18\lsim x_B\lsim 0.94$.
The best-fit parameters, along with the $\chi^2$ per degree of freedom, are 
listed in Table~\ref{tabx}.
\begin{table}[ht]
\begin{center}
\begin{tabular}{||c|c||}\hline
$\alpha$&$0.90\pm 0.15$  \\
\hline
$\beta$&$16.23\pm 1.37$  \\
 \hline
$\chi^2(\alpha,\beta)$/dof&33.42/31 \\
\hline
$\delta$&$17.07\pm 0.39 $ \\
\hline
$\chi^2(\delta)$/dof&33.80/32  \\
\hline
$\epsilon$&$(1.71\pm 0.09)\times 10^{-3}$\\
\hline
$\chi^2(\epsilon)$/dof&166.36/32\\
 \hline
\end{tabular}
\end{center}
\caption{\label{tabx}
Results of combined
fits to $e^+e^-\to b\bar b$ data from the ALEPH and SLD Collaborations,
using NLO coefficient functions,
NLL DGLAP evolution and NLL soft-gluon
resummation.
I have set $\Lambda=200$~MeV, $\mu_{0F}=\mu_0=m_b=5$~GeV and
$\mu_F=\mu=\sqrt{s}=91.2$~GeV.}
\end{table}
\par
\begin{figure}
\centerline{\psfig{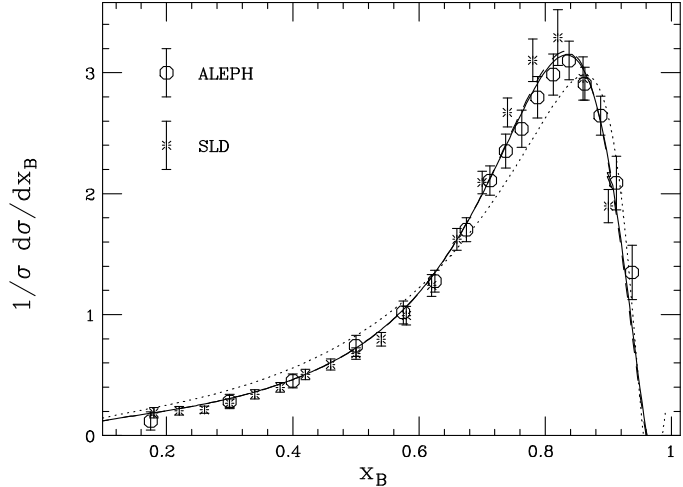}}
\caption{ALEPH and SLD data on $B$-hadrons, along with the best-fit curves
according to the power law (\ref{ab}) (solid line), the
Kartvelishvili model (dashes) and the Peterson model (dots), 
with $\alpha$, $\beta$, $\delta$ and $\epsilon$ given by the
central values reported in Table~\ref{tabx}.
All curves include NLL collinear and soft resummation in the parton-level
calculation.}
\label{fitall}
\end{figure}

 The power law (\ref{ab}) and the Kartvelishvili model (\ref{kk}) yield
very good fits, while the Peterson model (\ref{peter}) 
is unable to reproduce the data.
In \cite{cor} it was also shown that the inclusion of NLL soft resummation in 
the coefficient function and in the initial condition of the perturbative
fragmentation function is necessary to reproduce the data.
Figure~\ref{fitall} shows the data samples, along with the fitting curves 
corresponding to the central values of $\alpha$, $\beta$, $\delta$ and 
$\epsilon$ quoted in Table~\ref{tabx}.
We see that models (\ref{ab}) and (\ref{kk}) yield approximately the
same distribution: in fact, from Table~\ref{tabx} we learn that,
within the error ranges, $\alpha$ is consistent 
with 1 and $\beta$ with $\delta$. The Peterson non-perturbative fragmentation
function is unable to fit the data.
\par
\begin{figure}
\centerline{\psfig{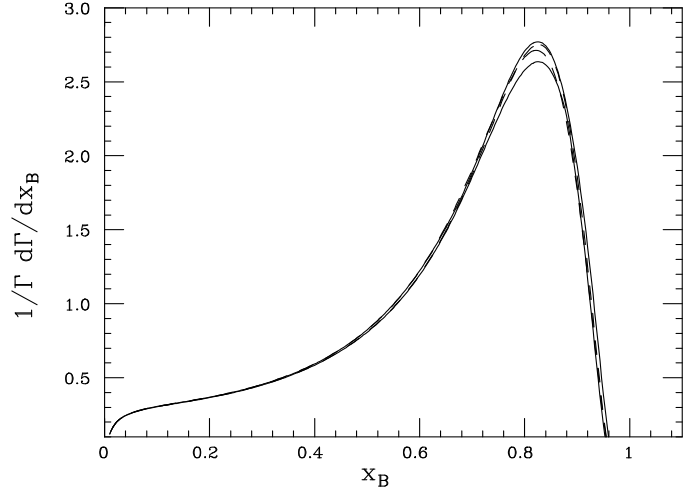}}
\caption{$B$-hadron spectrum in Higgs decay: the hadronization is described
according to the power law (\ref{ab}) (solid) and the Kartvelishvili
model (\ref{kk}) (dashes). Plotted are one-standard-deviation bands
for the non-perturbative parameters $\alpha$, $\beta$ and $\delta$.}
\label{hadh}
\end{figure}

Given the results of Table~\ref{tabx}, we can predict the $B$-hadron
spectrum in Higgs or top decay. In Fig.~\ref{hadh} we show the $x_B$ 
distribution in Higgs decay according to models (\ref{ab}) and (\ref{kk}).
Plotted are the edges of bands at one-standard-deviation confidence level.
We can see that the predictions yielded by the two models are in statistical 
agreement.
\par
\begin{figure}
\centerline{\psfig{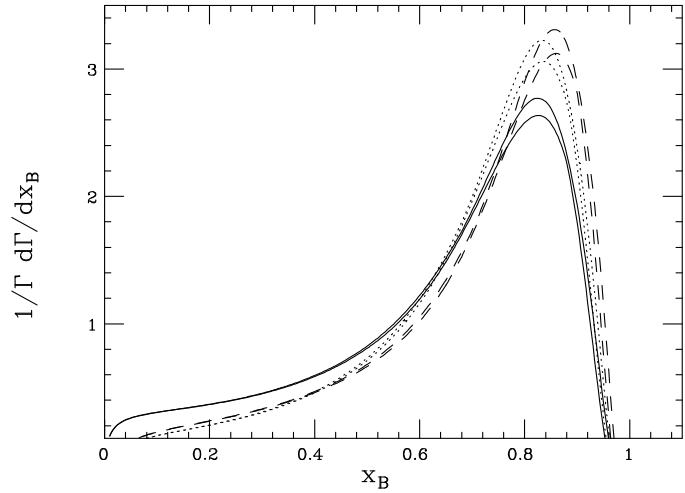}}
\caption{One-standard-deviation prediction for 
$B$-hadron spectra in Higgs decay (solid), top decay (dashes)
and $e^+e^-$ annihilation at $\sqrt{s}=91.2$~GeV (dotted),
according to the hadronization model (\ref{ab}).}
\label{hadhzt}
\end{figure}

In Fig.~\ref{hadhzt} we compare $b$-flavoured hadron
spectra in $e^+e^-$ annihilation
at LEP energy, Higgs and top decay, according to the power law (\ref{ab}). 
The results are
similar to the ones at parton level presented in Fig.~\ref{hzt}: the top-decay
spectrum is shifted toward larger $x_B$, the $H\to b\bar b$ one is the 
highest at low $x_B$, the $e^+e^-$-prediction annihilation lies 
between the two.
\par Finally, we can use the moments on $B$ production in $e^+e^-$ annihilation
from the DELPHI Collaboration \cite{delphi} to predict the same moments
in top and Higgs decay. 
In $N$-space we have: $\sigma_N^B=\sigma_n^b D^{np}_N$,
$\Gamma_N^B=\Gamma^b_N D^{np}_N =\Gamma^B_N\sigma_N^B/\sigma_N^b$,
where $\sigma_N$ and $\Gamma_N$ are the moments of 
$e^+e^-$-annihilation cross section, Higgs or top width, 
at hadron or parton level.
The results are shown in Table~\ref{tabn}: the moments of the top-decay
width are the largest, the ones of the Higgs width are the lowest, and 
the $e^+e^-$-annihilation ones are in the middle.
This result is consistent with what is observed in $x_B$-space in 
Fig.~\ref{hadhzt}.
\begin{table}[ht]
\begin{tabular}{| c | c c c c |}
\hline
& $\langle x\rangle$ & $\langle x^2\rangle$ & $\langle x^3\rangle$
& $\langle x^4\rangle$ \\
\hline
$e^+e^-$ data $\sigma_N^B$&0.7153$\pm$0.0052 &0.5401$\pm$0.0064 &
0.4236$\pm$0.0065 &0.3406$\pm$0.0064  \\
\hline
\hline
$e^+e^-$ NLL $\sigma_N^b$ & 0.7801 & 0.6436 & 0.5479 & 0.4755  \\
\hline
$D^{np}_N$ [B] & 0.9169 & 0.8392 & 0.7731 & 0.7163 \\
\hline
\hline
$t$-decay NLL $\Gamma^b_N$ & 0.7884 & 0.6617 & 0.5737 & 0.5072 \\
\hline
$t$-decay $\Gamma^B_N$ & 0.7228 & 0.5553 & 0.4435 & 0.3633 \\
\hline
$H$-decay NLL $\Gamma^b_N$ & 0.7578 & 0.6162 & 0.5193 & 0.4473  \\
\hline
$H$-decay $\Gamma^B_N$ & 0.6948 & 0.5171 & 0.4015 & 0.3204 \\
\hline
\end{tabular}
\caption{\label{tabn}\small  Experimental data for the moments
$\sigma^B_N$ from
DELPHI~\protect\cite{delphi}, the resummed $e^+e^-$ perturbative
calculations for $\sigma^b_N$~\protect\cite{cc},
the extracted non-perturbative contribution
$D^{np}_N$. Using the resummed 
perturbative result $\Gamma_N^b$ on top and Higgs
partonic widths, a prediction for
the moments $\Gamma^B_N$ $t\to bW$ and $H\to b\bar b$ processes is given.}
\end{table}
\par In summary, I have reviewed recent results on bottom-quark fragmentation
in $e^+e^-$ annihilation, Higgs and top decay. Parton-level spectra
exhibit a relevant impact of the inclusion of NLL collinear and soft 
resummation; using information from LEP and SLD experiments,
$b$-flavoured hadron energy distributions have been predicted.
\section*{Acknowledgements}
The results on top-quark decay were obtained in collaboration with
M. Cacciari and A.D. Mitov.

\end{document}